\documentclass[twocolumn,aps,pra,showpacs,superscriptaddress]{revtex4}

\usepackage{graphicx}
\usepackage{amsmath,amssymb}
\usepackage{color}
\def\endproof{\vrule height6pt width6pt depth0pt} 

\begin{document}


\title{Simple Explanation of the Quantum Violation of a Fundamental Inequality}


\author{Ad\'an Cabello}
\email{adan@us.es}
\affiliation{Departamento de F\'{\i}sica Aplicada II, Universidad de Sevilla, E-41012 Sevilla, Spain}


\date{\today}



\begin{abstract}
We show that the maximum quantum violation of the Klyachko-Can-Binicio\u{g}lu-Shumovsky (KCBS) inequality is {\em exactly} the maximum value satisfying the following principle: The sum of probabilities of pairwise exclusive events cannot exceed~1. We call this principle ``global exclusivity,'' since its power shows up when it is applied to global events resulting from enlarged scenarios in which the events in the inequality are considered jointly with other events. We identify scenarios in which this principle singles out quantum contextuality, and show that a recent proof excluding nonlocal boxes follows from the maximum violation imposed by this principle to the KCBS inequality.
\end{abstract}



\maketitle


{\em Introduction.---}Quantum mechanics (QM) cannot be explained neither with noncontextual hidden variable (NCHV) theories \cite{Specker60, Bell66, KS67} nor with local hidden variable theories \cite{Bell64}. It is in this sense in which QM is ``contextual'' and ``nonlocal.'' In the quest for the ``shocking principle'' \cite{Fuchs11} behind QM, it has been argued \cite{Fuchs11, Cabello11, Cabello12} that, rather than looking for this principle in the answer to the question of why QM is not more nonlocal, as pursued in Refs.~\cite{PR94, V05, PPKSWZ09, NW09, OW10}, one should start by answering another question: Why is QM not more contextual? The reasons behind this twist are both aesthetical, since contextuality is a generalization of nonlocality that does not privilege spacelike separated tests (which do not seem to play any special role in the axioms of QM), and practical, since characterizing the maximum quantum contextuality for a given graph of relationships of exclusivity is simple (it is the solution of a single semidefinite program \cite{CSW10}), while characterizing the maximum quantum nonlocality of a given Bell inequality is much more difficult (it is the solution of a converging hierarchy of semidefinite programs \cite{NPA07,NPA08,PV10}). Moreover, the observation that no principle based on bipartite information concepts can single out quantum correlations \cite{GWAN11} stimulates the search for principles that do not attribute any fundamental role to the ability to distinguish parties.

Quantum correlations are contextual in the sense that they cannot be explained assuming that the result of a test $A$ is independent of whether $A$ is performed together with a compatible test $B$ or with a compatible test $C$ (which may be incompatible with $B$). This is the assumption of noncontextuality (NC) of results, and NCHV theories are those making this assumption. Two tests are compatible when, for any preparation, each test always yields identical result, no matter how many times the tests are performed or in which order. Contextuality is revealed by the violation of NC inequalities \cite{CFHR08, KCBS08, Cabello08, SBKTP09, BBCP09}, which are restrictions satisfied by any NCHV theory. Bell inequalities are a particular type of NC inequalities in which the tests are not only compatible but spacelike separated.

In this Letter we show that a simple principle explains the maximum quantum violation of a fundamental NC inequality. Indeed, simple applications of this principle explains nature's maximum contextuality (assumed to be given by QM) in many other scenarios.

Two events are exclusive if they cannot be simultaneously true. By $a,b,\ldots,c|x,y,\ldots,z$ we denote the event ``the results $a,b,\ldots,c$ are respectively obtained when the compatible tests $x,y,\ldots, z$ are performed.'' Two events $a,b,\ldots,c|x,y,\ldots,z$ and $a',b',\ldots,c'|x',y',\ldots,z'$ are exclusive if $x=x'$ and $a \neq a'$, or if $y=y'$ and $b \neq b'$,\ldots, or if $z=z'$ and $c \neq c'$.

The principle is that the sum of the probabilities of pairwise exclusive events cannot exceed~1. This principle follows from Specker's observation that pairwise decidable events must not necessarily be jointly decidable \cite{Specker60}, and from Boole's axiom of probability stating that the sum of the probabilities of events that are jointly exclusive cannot exceed~1 \cite{Boole62}. Specker conjectured that ``the fundamental theorem of QM'' might be that ``if you have several questions and you can answer any two of them [i.e., if the corresponding propositions (or events) are pairwise decidable], then you can also answer all of them [i.e., the corresponding propositions are simultaneously (or jointly) decidable]'' \cite{Specker09}. This principle was used by Wright \cite{Wright78} to show that simple sets of events allow probabilities such that their sum can exceed the maximum classical value. In \cite{CSW10}, it was shown that quantum contextual and nonlocal correlations are bounded by this principle, and that the maximum value, satisfying this principle, of a sum of probabilities of events is given by the fractional packing number of the graph in which exclusive events are represented by adjacent vertices. It was also shown that, in QM, this maximum is upper bounded by the Lov\'asz number of this graph (see Appendix~\ref{App1}). This shows that this principle singles out all quantum correlations represented by a graph such that its Lov\'asz number equals its fractional packing number and such that the maximum quantum value reaches the Lov\'asz number. Therefore, this condition singles out the quantum correlations in the case of Bell inequalities for Greenberger-Horne-Zeilinger states \cite{Mermin90} and graph states \cite{GTHB05, CGR08, CPSS12}, some bipartite Bell inequalities \cite{Cabello01, AGACVMC12}, and all the state-independent NC inequalities in \cite{Cabello08}. The results in \cite{CSW10} have been used for identifying new quantum correlations \cite{ADLPBC12, Cabello12b, DHANBSC12} and quantum advantages \cite{NDSC12} singled out by this condition. A more recent work \cite{A12} obtains tighter bounds on nonlocal correlations by applying that the sum of probabilities of pairwise exclusive events cannot exceed 1, to events in which tests $x,y,\ldots,z$ are pairwise spatially separated.

As we will see, the power of the principle to single out physical limits increases when it is applied to global events in which the events in the NC inequality are considered jointly with other events. Different families of global events will be described below. For this reason, hereafter we will refer to it as ``global exclusivity'' (GE), while we will use ``exclusivity'' (E) when it is applied to the original events in the NC inequality.


\begin{figure}[t]
\centerline{\includegraphics[width=0.48\columnwidth]{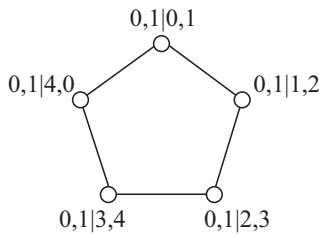}}
\caption{\label{Fig1}Graph of the relationships of exclusivity between the 5 events in the KCBS inequality. Vertex $a,b|x,y$ represents the event ``the results $a$ and $b$ are respectively obtained when compatible tests $x$ and $y$ are performed.'' Exclusive events are represented by adjacent vertices.}
\end{figure}


{\em Maximum quantum violation of the KCBS inequality.---}The simplest physical system violating a NC inequality is a qutrit (i.e., a three-level quantum system). The simplest NC inequality violated by a qutrit is the Klyachko-Can-Binicio\u{g}lu-Shumovsky (KCBS) inequality \cite{KCBS08}, which is necessary and (together with other NC inequalities) sufficient for noncontextuality \cite{KCBS08, AQBTC12}. Its quantum violation is behind the quantum violation of other NC inequalities \cite{SBBC11}. All this makes the KCBS inequality fundamentally important in QM. Its quantum violation has been recently tested with photons \cite{LLSLRWZ11, AACB12}, and can be used to put lower bounds to the quantum dimension of physical systems \cite{GBCKL12}.

If we complete the KCBS inequality with its maximum quantum violation and the upper bound imposed by E, we obtain the following expression:
\begin{equation}
 \sum_{i=0}^4 P(0,1|i,i+1) \stackrel{\mbox{\tiny{ NCHV}}}{\leq} 2 \stackrel{\mbox{\tiny{ QM}}}{\leq} \sqrt{5} \stackrel{\mbox{\tiny{ E}}}{\leq} \frac{5}{2},
 \label{KCBS}
\end{equation}
where $P(0,1|i,i+1)$ denotes the probability of the event $0,1|i,i+1$, the sum is taken modulo 5, $\stackrel{\mbox{\tiny{ NCHV}}}{\leq} 2$ indicates that $2$ is the maximum value for NCHV theories \cite{CSW10, Wright78, KCBS08}, $\stackrel{\mbox{\tiny{ QM}}}{\leq} \sqrt{5}$ indicates that $\sqrt{5} \approx 2.236$ is the maximum value in QM (even for systems of arbitrary dimension) \cite{CSW10, KCBS08, BBCGL11}, and $\stackrel{\mbox{\tiny{ E}}}{\leq} \frac{5}{2}$ indicates that $\frac{5}{2}$ is the maximum value in any theory in which the sum of the probabilities of events that are pairwise exclusive cannot exceed 1, applied to the 5 events $0,1|i,i+1$. This limit was found by Wright \cite{Wright78} and rediscovered in \cite{CSW10}.

The question is, why does the quantum violation of the KCBS inequality stop at $\sqrt{5}$ \cite{Cabello12b}.


{\em Result 1}: $\sqrt{5}$ is the maximum violation of the KCBS inequality allowed by GE.

\begin{figure}[t]
\centerline{\includegraphics[width=\columnwidth]{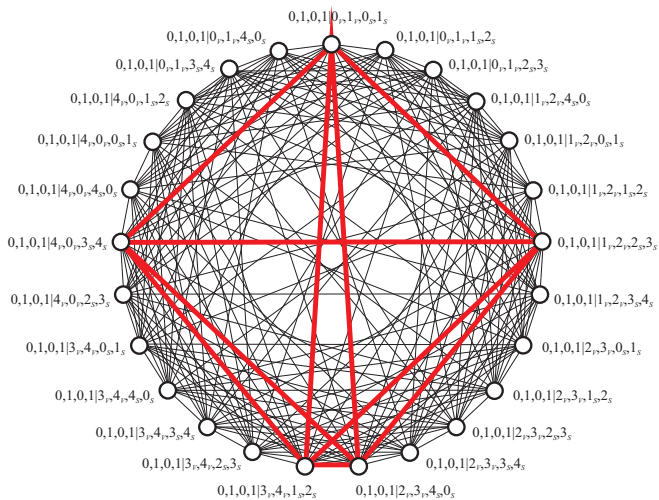}}
\caption{\label{Fig1b} (Color online) Graph of the relationships of exclusivity between the 25 events obtained from considering two experiments testing the KCBS inequality. Each vertex represents an event $a,b,c,d|x_V,y_V,z_S,k_S$ denoting
``the results $a$ and $b$ are respectively obtained when compatible tests $x$ and $y$ are performed in Vienna, and the results $c$ and $d$ are respectively obtained when compatible tests $z$ and $k$ are performed in Stockholm.'' Exclusive events are represented by adjacent vertices. Any event belongs to a set of 5 pairwise exclusive events. One of these sets is indicated in red.}
\end{figure}

{\em Proof.---}The graph of the relationships of exclusivity between the 5 events tested in the KCBS inequality is shown in Fig.~\ref{Fig1}. Consider two independent experiments testing the KCBS inequality, one performed in Vienna on a system prepared in a quantum pure state \cite{LLSLRWZ11} and another one in Stockholm on a different system also prepared in a quantum pure state \cite{AACB12}. The two experiments might even be spacelike separated. There are two types of ``local'' events: the 5 local events $0,1|i_V, i+1_V$, with $i=0,\ldots,4$, corresponding to the Vienna experiment, and the 5 local events $0,1|j_S, j+1_S$, with $j=0,\ldots,4$, corresponding to the Stockholm experiment. From them we can construct the 25 ``global'' events $0,1,0,1|i_V, i+1_V, j_S, j+1_S$. If we draw the graph of the relationships of exclusivity between these 25 global events, we obtain the graph in Fig.~\ref{Fig1b}. The important point in this graph is to notice that, for any $i,j \in \{0,\ldots,4\}$ and taking the sum modulo 5, the 5 events
\begin{subequations}
\begin{align}
 0,1,0,1|&i_V,i+1_V,j_S,j+1_S,\\
 0,1,0,1|&i+1_V,i+2_V,j+2_S,j+3_S,\\
 0,1,0,1|&i+2_V,i+3_V,j-1_S,j_S,\\
 0,1,0,1|&i+3_V,i+4_V,j+1_S,j+2_S,\\
 0,1,0,1|&i-1_V,i_V,j+3_S,j+4_S
\end{align}
\end{subequations}
{\em are pairwise exclusive}. Therefore, according to E, the sum of their probabilities cannot exceed~1. The maximum value, satisfying E, of the sum of the probabilities of the 25 global events is 5, and the only way to reach it is by assigning probability $\frac{1}{5}$ to each and every one of the global events.


$a,b|i_V, i+1_V$ and $c,d|j_S, j+1_S$ are two completely independent events. Since the joint probability of two independent events is the product of their probabilities, then $P(a,b,c,d|i_V, i+1_V, j_S, j+1_S)=P(a,b|i_V, i+1_V) P(c,d|j_S, j+1_S)$. Therefore, the maximum probability satisfying GE for the local events corresponding to the Vienna and Stockholm experiments is $\frac{1}{\sqrt{5}}$. The sum of the 5 probabilities of the local events corresponding to the Vienna (or Stockholm) experiment gives $\sqrt{5}$, which is exactly the maximum quantum violation of the KCBS inequality. \hfill \endproof

An alternative proof is provided in Appendix~\ref{App2}.

The graph representing the relationships of exclusivity of the global events obtained from two copies of an experiment whose relationships of exclusivity are represented, in both cases, by the same graph $G$, corresponds to the OR product (also called co-normal product, disjunctive product, or disjunction product) of two copies of $G$ \cite{HIK11}. The OR product of two graphs $G$ and $H$ is a new graph $G \ast H$ whose vertex set is $V(G)\times V(H)$ and in which two vertices $(g,h)$ and $(g',h')$ in $G \ast H$ are adjacent if $g$ and $g'$ were adjacent vertices in $G$ or $h$ and $h'$ were adjacent vertices in $H$. The graph in Fig.~\ref{Fig1b} is the OR product of two copies of the graph in Fig.~\ref{Fig1}.
Similarly, the graph of the relationships of exclusivity for $n$ copies of the experiment is given by the OR product of $n$ copies of $G$, denoted as $G^{\ast n}$.


\begin{figure}[t]
\centerline{\includegraphics[width=0.9\columnwidth]{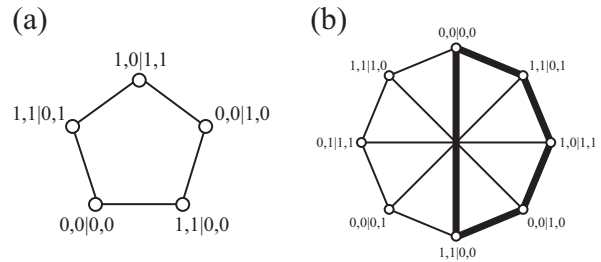}}
\caption{\label{Fig2} (a) Graph of the relationships of exclusivity between 5 events to which a PR box assigns a probability $\frac{1}{2}$ to each event. (b) Graph of the relationships of exclusivity between the 8 events involved in the CHSH inequality. This graph is the 8-vertex (1,4)-circulant graph $Ci_8(1,4)$. It contains 8 induced pentagons. The pentagon emphasized corresponds to the one in (a). Vertex $ab|xy$ represents the event ``the results $a$ and $b$ are respectively obtained when spacelike separated tests $x$ and $y$ are performed,'' where $x$ is performed in Alice's side and $y$ in Bob's. Exclusive events are represented by adjacent vertices.}
\end{figure}


{\em The CHSH inequality.---}The Clauser-Horne-Shimony-Holt (CHSH) inequality \cite{CHSH69} is the tight Bell inequality corresponding to the bipartite scenario in which Alice chooses between two tests $x \in \{0,1\}$ and Bob chooses between two tests $y \in \{0,1\}$. Each test has two possible results: Alice's are denoted $a \in \{0,1\}$ and Bob's $b \in \{0,1\}$. If we complete the CHSH inequality with its maximum violation in QM (Tsirelson's bound \cite{Cirelson80}) and the upper bound imposed by nonsignaling \cite{PR94} (which equals the one imposed by E), we obtain the following expression:
\begin{equation}
 \sum P(a,b|x,y) \stackrel{\mbox{\tiny{ NCHV,LHV}}}{\leq} 3 \stackrel{\mbox{\tiny{ QM}}}{\leq} 2 + \sqrt{2} \stackrel{\mbox{\tiny{ E,NS}}}{\leq} 4,
 \label{CHSH}
\end{equation}
where the sum is extended to all $x,y\in \{0,1\}$ and $a,b\in \{0,1\}$ such that $a \oplus b=xy$, where $\oplus$ denotes sum modulo 2, LHV means local hidden variables, and NS means nonsignaling.

A PR box \cite{PR94} is a two-party nonsignaling device which achieves the maximum algebraic violation of the CHSH inequality, which is equal to the maximum violation satisfying NS and E. A PR box produces joint probabilities $P(a,b|x,y)=\frac{1}{2}$, if $a \oplus b=xy$, and $0$ otherwise.

Now the question is, why are PR boxes not allowed in nature despite that they do not violate nonsignaling \cite{PR94}.

Many reasons have been given for this \cite{V05, PPKSWZ09, NW09, OW10, PW12}. Here we show that a recent proof \cite{A12} can be simplified by the following observation.


{\em Observation 1}: A PR box assigns probability $\frac{1}{2}$ to 5 joint probabilities of events whose relationships of exclusivity are exactly the ones of the KCBS inequality. Therefore, the proof of Result~1 does not only excludes Wright's assignment to the KCBS inequality, but also excludes PR's assignment to the CHSH inequality.


{\em Proof.---}A PR box assigns probability $\frac{1}{2}$ to the 5~joint probabilities of events whose relationships of exclusivity are represented in Fig.~\ref{Fig2} (a). \hfill \endproof


The proof of Result 1 does not single out Tsirelson's bound $2 + \sqrt{2} \approx 3.4142$, but states that the maximum quantum nonlocality should be less than or equal to $\frac{8}{\sqrt{5}} \approx 3.5778$. Interestingly, this is the same bound obtained by considering {\em all} the restrictions that E imposes, after assuming the principle of local orthogonality, to all possible combinations of the 64 events resulting from two PR boxes \cite{A12}. This emphasizes the fundamental role of the elementary Bell inequalities introduced in \cite{SBBC11} to understand the quantum violation of Bell inequalities.

The next question is, why does the quantum violation of the CHSH inequality stop at Tsirelson's bound \cite{PR94}.

Here we show that, if we only use GE applied to multiple copies of the CHSH experiment, then the answer is not known and is related to an open problem in graph theory. However, a curious observation can be made.


{\em Observation 2}: {\em If} the Shannon capacity \cite{Shannon56} of the 8-vertex (1,2)-circulant graph $Ci_8(1,2)$ were equal to its Lov\'asz number \cite{Lovasz79}, then GE applied to an infinite number of copies would single out Tsirelson's bound of the CHSH inequality.


{\em Proof.---}The relationships of exclusivity of the 8 events in the CHSH inequality are represented by the graph of Fig.~\ref{Fig2} (b), which is the 8-vertex (1,4)-circulant graph, $Ci_8(1,4)$, and is isomorphic to the 4-M\"{o}bius ladder, $M_4$ \cite{GH67}, and to the Wagner graph \cite{Wagner37}. It can be shown that, for $M_4$ (and similarly for vertex-transitive graphs; see Appendix~\ref{App2}), the maximum value allowed by GE for $n$ copies is given by the number of vertices of $M_4$ times the maximum probability $p_n$ that can be assigned to each and every local event without violating GE (applied to $n$ copies). This number is the inverse of the $n$th root of the clique number of the OR product of $n$ copies of $M_4$, denoted as $p_n (M_4)= \left[\omega(M_4^{\ast n})\right]^{-\frac{1}{n}}$, which, if $n \rightarrow \infty$, is exactly equal to $p_{\infty} (M_4)=\left[\Theta(\bar{M_4})\right]^{-1}$, where $\Theta(G)$ is the Shannon capacity of $G$. This correspondence can be seen by taking into account the following equalities: (i) $\omega(G)=\alpha(\bar{G})$, where $\bar{G}$ is the complement of $G$ and $\alpha(G)$ is the independence number of $G$ \cite{Diestel10}, (ii) $G^{\ast n}= \overline{\bar{G}^{\boxtimes n}}$, where $G^{\boxtimes n}$ is the strong product of $n$ copies of $G$ \cite{HIK11}, and (iii) $\Theta(G)=\lim_{n \rightarrow \infty} \left[\omega(G^{\boxtimes n})\right]^{\frac{1}{n}}$ \cite{Shannon56}.

The complement of $M_4$ is the 8-vertex (1,2)-circulant graph $Ci_8(1,2)$. Unfortunately, $Ci_8(1,2)$ does not belong to any of the classes of graphs for which the Shannon capacity is known \cite{Lovasz79}. However, if $\Theta(Ci_8(1,2))$ were equal to a well-known upper bound, its Lov\'asz number \cite{Lovasz79}, which is $\vartheta(Ci_8(1,2))=8-4 \sqrt{2}$, then GE (applied to an infinite number of copies) would exactly single out nature's nonlocality for the CHSH scenario. \hfill \endproof


Similar considerations can be found in \cite{A12}. Note that, even if $\Theta(Ci_8(1,2)) \neq \vartheta(Ci_8(1,2))$, this would not mean that GE is incapable of singling out Tsirelson's bound when applied to a different type of global events.


{\em Other scenarios where GE singles out quantum contextuality.---}Using previous results in graph theory and quantum contextuality, we can see that GE also singles out the maximum quantum contextuality for an infinite family of new scenarios.


{\em Result 2}: Any self-complementary vertex-transitive graph with $n$ vertices such that $n \neq p^2$ with $p$ prime, corresponds to a scenario in which GE applied to two copies singles out the maximum quantum contextuality.


{\em Proof.---}A well-known result in graph theory states that for all self-complementary vertex-transitive graphs with $n$ vertices, (i) $\Theta(G)=\vartheta(G)=\sqrt{n}$ and $p_2 (G)= \left[\omega(G \ast G)\right]^{-\frac{1}{2}}=\left[\Theta(G)\right]^{-1}$ \cite{Lovasz79}. On the other hand, NC inequalities violated by QM are represented by graphs such that (ii) $\alpha(G) < \vartheta(G)$ \cite{CSW10}. Since for any graph $G$, $\alpha(G) \le \vartheta(G)$ and $\alpha(G)$ is, by definition, an integer number, the self-complementary vertex-transitive graphs with $n \neq p^2$ and $p$ prime satisfy simultaneously (i) and (ii), since, for them, $\vartheta(G)=\sqrt{n}$ is not an integer number. \hfill \endproof


The simplest member of this family of graphs is the pentagon $C_5$ corresponding to the KCBS inequality. Other members are $Ci_{13}(1,2,6)$, $Ci_{13}(1,3,4)$ (or Paley-13), $C_{17}(1,2,3,6)$, $C_{17}(1,2,4,8)$ (or Paley-17), and $C_{17}(1,3,4,5)$.


{\em Conclusions and conjecture.---}We have provided an extremely simple answer to the question ``What physical principle limits quantum contextuality in the scenario of the KCBS inequality?'' \cite{Cabello12b}. The answer is: GE, namely that the sum of the probabilities of pairwise exclusive events cannot exceed~1, applied to an extended set of events comprising not only the events in the KCBS inequality itself, but also other events that may be tested simultaneously. In addition, we have shown that GE applied to one or two copies singles out the maximum quantum contextuality of a family of NC inequalities (Results 1 and 2) and the maximum quantum nonlocality of some Bell inequalities \cite{Mermin90, GTHB05, CGR08, CPSS12, Cabello01, AGACVMC12}.

A simple application of GE automatically excludes PR nonlocal boxes (as also shown in \cite{A12}) and the connection between the KCBS and the CHSH inequalities \cite{SBBC11} provides a much simpler proof than anyone proposed before \cite{V05, PPKSWZ09, NW09, OW10, PW12}.

It is still unclear whether GE can single out nature's maximum contextuality for any graph. It seems to be very likely that GE applied to other types of global events will single out quantum contextuality for a larger family of scenarios. Future work should explore the power of GE applied to other families of global events: for example, (i) to global events arising from considering ancillary experiments with maximum contextuality already constrained by E, (ii) to graphs in which the weights of the events are different than those in the original NC inequality, (iii) to complement graphs, (iv) to scenarios in which additional compatible measurements are considered, and (v) to combinations of (i)--(iv). If, for any graph $G$, we can produce a larger graph $G'$ such that the fractional packing number of $G'$ induces values on the probabilities of the events in $G$ such that the sum of them is the Lov\'asz number of $G$, then we would prove that GE singles out nature's maximum contextuality for any graph, providing a surprising and extraordinarily simple answer to the question of why QM is not more contextual. In any case, it is remarkable that such a simple principle allows us to single out quantum contextuality in many scenarios.


\begin{acknowledgments}
This work was supported by the Project No.\ FIS2011-29400 (Spain). The author thanks A.~Ac\'{\i}n, I.~Bengtsson, C.~Budroni, S.~Klav\v{z}ar, A.~J.~L\'opez-Tarrida, J.~Henson, S.~Kochen, J.~R.~Portillo, A.~B.~Sainz, and S.~Severini for helpful discussions.
\end{acknowledgments}


\appendix


\section{Previous results}
\label{App1}


In Ref.~\cite{CSW10} it is shown that every NC inequality can be associated to a graph $G$ such that the NCHV, QM, and E bounds are related to three characteristic numbers of $G$. This graph is constructed by reexpressing the linear combination of joint probabilities of
events in the NC inequality as a sum $S$ of joint probabilities of events (plus some constant). Then, $G$ is the graph in which each of the events in $S$ is represented by a vertex and exclusive events are represented by adjacent vertices.

The upper bound of $S$ for NCHV theories is exactly given by the independence number of $G$, $\alpha(G)$, which is the maximum number of pairwise nonadjacent vertices of $G$ \cite{Diestel10}. This bound is always saturated by some NCHV theory (and by some LHV theory, when the NC inequality is a Bell inequality). Computing $\alpha(G)$ is NP-complete.

The upper bound of $S$ for QM is given by the Lov\'asz number of $G$, $\vartheta(G)$, which is $\max \sum_{i \in V(G)} |\langle\psi|v_{i}\rangle|^{2}$, where $V(G)$ is the set of vertices of $G$, and the maximum is taken over all unit vectors $|\psi\rangle$ and $|v_{i}\rangle$ and all dimensions, where $\{|v_{i}\rangle\}$ is an orthogonal representation of $G$ (which means that each vertex is assigned a vector so that adjacent vertices are assigned orthogonal vectors) \cite{Lovasz79}. Interestingly, $\vartheta(G)$ can be computed to any desired precision in polynomial time \cite{GLS81}.

Finally, the upper bound of $S$ satisfying E is exactly given by the fractional packing number of $G$, $\alpha^* (G)$, which is $\max \sum_{i\in V(G)} w_i$, where the maximum is taken over all $0 \leq w_i\leq 1$ and for all cliques $C_j$ (subsets of pairwise linked vertices) of $G$, under the restriction $\sum_{i \in C_j} w_i \leq 1$ \cite{SU97}.


\section{Alternative proof of Result 1}
\label{App2}


If $G$ is vertex-transitive (VT) [i.e., given any two vertices $v_1$ and $v_2$ of $G$, there is some automorphism $f:V(G)\rightarrow V(G)$ such that $f(v_1)=v_2$], then the OR product $G \ast G$ is also VT. Then, the fractional packing number of a VT graph $G$ is its order $|G|$ (i.e., the number of vertices of $G$) times the probability assigned to each event in the case where the sum of all the probabilities is maximum (respecting E). In other words, for VT graphs,
\begin{equation}
 \alpha^*(G) \stackrel{\mbox{\tiny{ VT}}}{=} \frac{|G|}{\omega(G)},
 \label{vt}
\end{equation}
where $\omega(G)$ is the clique number of $G$, which is the maximum number of pairwise adjacent vertices.

For the graph of the relationships of exclusivity between global events resulting from considering two copies of $C_5$ (i.e., the graph in Fig.~2), we obtain that the maximum contextuality for the sum of the probabilities of the global events is given by
\begin{equation}
 \alpha^* (C_5 \ast C_5)= 5.
 \label{max}
\end{equation}
The only way to reach this value is by assigning probability $\frac{1}{5}$ to each and every one of the 25 global events.

Since the joint probability of independent events is the product of the probabilities of the local events, then the maximum \eqref{max} forces the probabilities of the local events $a,b|i_V, i+1_V$ and $c,d|j_S, j+1_S$ to be $\frac{1}{\sqrt{5}}$, which leads to the maximum quantum violation of the KCBS inequality. \hfill \endproof



\end{document}